\documentclass[10pt,twocolumn,twoside]{IEEEtran}

\usepackage{amsmath}
\usepackage{amsfonts}
\usepackage{amssymb}
\usepackage{mathrsfs}
\usepackage{color}
\usepackage{hyperref}
\usepackage{booktabs}
\usepackage{graphicx}
\usepackage{epstopdf}
\usepackage{epsfig}
\usepackage{cite}
\usepackage{enumerate}
\usepackage{amsopn}
\usepackage{ntheorem}
\usepackage{bm}
\usepackage{algorithm}
\usepackage{algpseudocode}

\newtheorem{lemma}{Lemma}
\newtheorem{theorem}{Theorem}
\newtheorem{remark}{Remark}
\newtheorem{corollary}{Corollary}

\newtheorem{example}{Example}

\title{Time Synchronization Attack and Countermeasure for Multi-System Scheduling in Remote Estimation}

\author{
	Ziyang Guo$^*$, Yuqing Ni$^*$, Wing Shing Wong$^\dag$, Ling Shi$^*$
	\thanks{
		$*$: Electronic and Computer Engineering, The Hong Kong University of Science and Technology, Clear Water Bay, Kowloon, Hong Kong (e-mail: {zguoae@connect.ust.hk}, {yniac@connect.ust.hk}, {eesling@ust.hk}).}

	\thanks{
		$\dag$: Department of Information Engineering, The Chinese University of Hong Kong, Shatin, Hong Kong (e-mail: {wswong@ie.cuhk.edu.hk}).}

	\thanks{
		The research was funded by Schneider Electric,  Lenovo Group (China) Limited and the Hong Kong Innovation and Technology Fund (ITS/066/17FP) under the HKUST-MIT Research Alliance Consortium.}
}

\DeclareMathOperator{\Tr}{Tr}
\DeclareMathOperator{\BB}{B\&B}

\begin{document} \maketitle
\begin{abstract}
We consider time synchronization attack against multi-system scheduling in a remote state estimation scenario where a number of sensors monitor different linear dynamical processes and schedule their transmissions through a shared collision channel. We show that by randomly injecting relative time offsets on the sensors, the malicious attacker is able to make the expected estimation error covariance of the overall system diverge without any system knowledge. For the case that the attacker has full system information, we propose an efficient algorithm to calculate the optimal attack, which spoofs the least number of sensors and leads to unbounded average estimation error covariance. To mitigate the attack consequence, we further propose a countermeasure by constructing shift invariant transmission policies and characterize the lower and upper bounds for system estimation performance. Simulation examples are provided to illustrate the obtained results.
\end{abstract}	

	
\section{Introduction}
Cyber-physical systems (CPS) refer to systems integrating sensing, computation, communication and control techniques with physical processes. With wide applications in different critical infrastructures such as smart grids, intelligent transportation and health monitoring system, CPS have attracted great research interest during the past decade~\cite{kim2012cyber}. In such systems, wireless sensors play an indispensable role due to advantages such as low cost, easy installation, self-power and inherent intelligent processing capability~\cite{gungor2009industrial}. However, new issues arise naturally with the widespread deployment.

Since most wireless sensors in practical applications are battery-powered, and replacing old batteries which run out of energy is usually costly or even impossible in some extreme environments, one critical issue is how to efficiently allocate the transmission power of the sensors. In~\cite{shi2011sensor}, the authors studied the scheduling problem of whether to send the sensory data to the remote estimator or not under transmission energy constraint and obtained the optimal off-line sensor schedule. In~\cite{ren2014dynamic}, the authors proposed an on-line sensor schedule based on the acknowledgment signal which contains the real-time data-dropout information. It was shown in~\cite{han2014online} that the on-line sensor schedule improves the estimation performance significantly compared with the off-line case with only a 1-bit feedback. Besides the above works focusing on single-sensor systems, multi-sensor and multi-system scheduling problems have received more attention recently~\cite{yang2015deterministic,han2017optimal,wu2017optimal}. Specifically, a sensor scheduling problem was investigated for a multi-sensor system under communication constraints in~\cite{yang2015deterministic} and the optimal sensor selection schemes were obtained for reliable and packet-dropping channels. The design of a collision-free transmission scheduling for multiple linear dynamical systems was studied in~\cite{han2017optimal} and an asymptotic periodic schedule was proved to be optimal. Moreover, the multi-system scheduling problem was extended to the scenario with packet drop and packet length constraint in~\cite{wu2017optimal}.

Due to the wireless communication modes and the complex interconnection between cyber information layer and physical components, communication channels are vulnerable to malicious attacks, which may raise security issues in CPS. Availability attacks, also known as Denial-of-Service attacks, which block the communication channels and prevent legitimate access to system components, were investigated for resource-constrained attacker in~\cite{qin2017optimal,li2015jamming}. Integrity attacks, another main category of cyber attacks, attempt to cripple the system functionality while remaining undetectable by intercepting and modifying the transmitted data packets. Different implementations of integrity attacks were studied, including replay attack~\cite{mo2009secure}, false-data injection attack~\cite{liu2011false,mo10data}, innovation-based deception attack~\cite{guo2017optimal,guo2018worst}, etc.

Note that most of the multi-system scheduling problems considered in the literature~\cite{yang2015deterministic,han2017optimal,wu2017optimal} assume that all the subsystems are working in a synchronous manner. Since Global Positioning System (GPS) signal is highly accurate, stable and free for timing, GPS-based measuring devices have been vastly deployed in critical infrastructures to guarantee the time synchronization between subsystems and thus achieve desirable performance~\cite{GPStiming}. However, malicious agents may disturb the time synchronization among sensors by introducing counterfeit GPS signals. Some existing works~\cite{tippenhauer2011requirements,wang2015time,zhang2013time} and real world GPS spoofing attacks~\cite{USspoofing} have shown the vulnerability of GPS signals and the possibility of spoofing the GPS receivers. Any successful time synchronization attack, which injects relative offsets on sensor clocks and desynchronizes the target system, may lead to a huge performance degradation. Motivated by these observations, we consider such an attack scenario in this work and analyze performance degradations for different attack information sets. To mitigate the attack effect, we further propose a defense strategy using transmission policies with shift invariance property. The main contributions of this paper are summarized as follows:
\begin{enumerate}
\item We consider time synchronization attack against multi-system scheduling and analyze the attack consequences for different scenarios. For the case that the attacker has no system knowledge, we show that it is possible to drive the expected estimation error covariance of the overall system to infinity (\textbf{Theorem~\ref{thm:existence_of_attack}} and \textbf{Corollary~\ref{cor:random-attack}}). For the case that the attacker has full system knowledge, we propose an efficient algorithm to calculate the optimal attack strategy that spoofs the least number of sensors and leads to unbounded estimation error covariance (\textbf{Theorem~\ref{thm:optimal-attack}} and \textbf{Algorithm~\ref{alg:optimal-attack}}).

\item To mitigate the attack consequence, we propose a countermeasure by constructing shift invariant transmission policies and derive the lower and upper bounds for the remote estimation error covariance when using the proposed countermeasure (\textbf{Theorem~\ref{thm:performance-bound}}). Moreover, we provide a procedure to construct the shift invariant transmission policies.
\end{enumerate}

The reminder of the paper is organized as follows. Section II introduces the system architecture and preliminaries about the multi-system scheduling. Section III considers time synchronization attack with or without system knowledge and analyzes the attack consequences. Section IV proposes a countermeasure by constructing shift invariant transmission policies and derives performance bounds of remote estimation error covariance. Simulation examples are provided in Section V. Some concluding remarks are made in the end.
	
\emph{Notations}: All vectors and matrices are named in boldface while scalars are not. $\mathbb{N}$ and $\mathbb{R}$ denote the sets of nonnegative integers and real numbers. $\mathbb{R}^n$ is the $n$-dimensional Euclidean space. $\mathbb{S}_{+}^{n}$ ($\mathbb{S}_{++}^{n}$) is the set of $n\times n$ positive semi-definite (definite) matrices. When $X\in \mathbb{S}_{+}^{n}$ ($\mathbb{S}_{++}^{n}$) , we simply write $X\succeq0$ ($X\succ0$). $\lfloor \cdot\rfloor$ means the floor function. For functions $f, f_1,f_2$, $f_1 \circ f_2$ is defined as $f_1 \circ f_2 (X)\triangleq f_1(f_2(X))$ and $f^k$ is defined as $f^k(X)\triangleq \underbrace{f\circ f\circ\cdots \circ f}_{k \ \textup{times}}(X)$ with $f^0(X)=X$.

\section{Preliminaries}
\subsection{System Model}
\begin{figure}[t]
	\centering
	\includegraphics[width=0.48\textwidth]{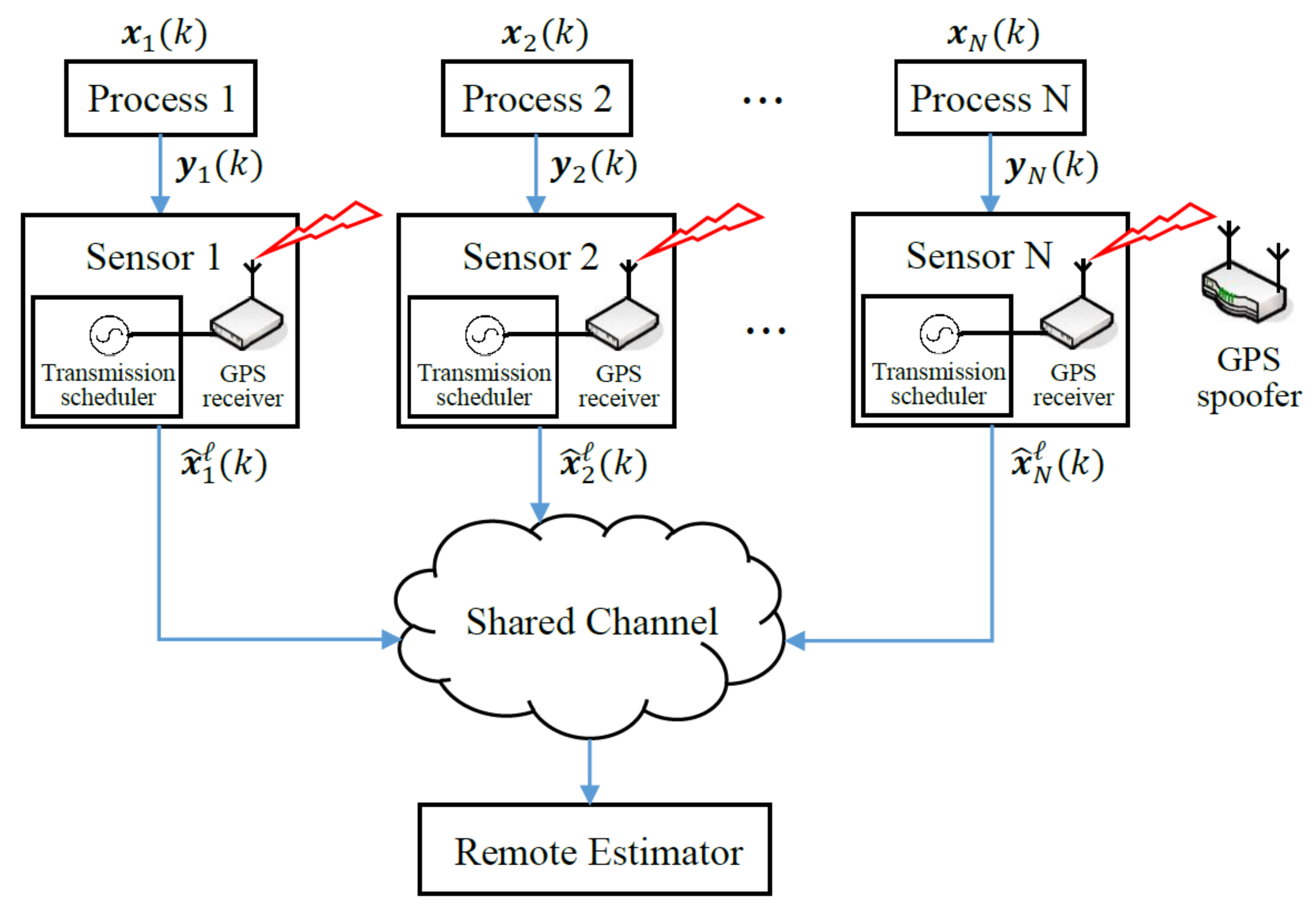}
	\caption{System block diagram.}
	\label{fig:system}
\end{figure}

Consider a system consisting of $N$ independent discrete linear time-invariant processes and $N$ sensors as depicted in Fig.~\ref{fig:system}. The $i$-th sensor monitors the $i$-th process:
\begin{align}
	\bm x_i(k+1)&=\bm A_i\bm x_i(k)+\bm w_i(k),\label{eqn:dynamic}\\
	\bm y_i(k)&=\bm C_i\bm x_i(k)+\bm v_i(k),\label{eqn:measurement}
\end{align}
where $i\in\mathcal{N}\triangleq\{1,2,...,N\}$, $k\in\mathbb{N}$ is the time index, $\bm x_i(k)\in\mathbb{R}^{n_i}$ is the state of the $i$-th process, $\bm y_i(k)\in\mathbb{R}^{m_i}$ is the measurement obtained by the $i$-th sensor. The process noise $\bm w_i(k)$, measurement noise $\bm v_i(k)$ and the initial state $\bm x_i(0)$ are mutually independent zero-mean Gaussian random variables with covariance $\bm Q_i\succeq \bm 0$, $\bm R_i\succ\bm 0$ and $\bm \Pi_i\succeq\bm 0$, respectively. To avoid trivial problem, all the processes are assumed to be unstable. The pair $(\bm A_i,\bm C_i)$ is detectable and $(\bm A_i,\sqrt{\bm Q_i})$ is stabilizable.

\subsection{Smart Sensor}
With the development of manufacturing techniques, many modern sensors are able to provide extra functions beyond those necessary for generating the measured quantity. The functions embedded might be signal processing, decision-making and anomaly alarm, which may promote the system performance~\cite{frank2013understanding}. In this work, each sensor is assumed to be smart in the sense that it is able to measure the state of its corresponding process and run the following Kalman filter to generate a local estimate:
\begin{align*}
	 \hat {\bm x}_i^{\ell-}(k)&=\bm A_i\hat {\bm x}_i^\ell(k-1),\\
	 \bm P_i^{\ell-}(k)&=\bm A_i\bm P_i^\ell(k-1)\bm A_i^\top + \bm Q_i,\\
	 \bm K_i^\ell(k)&=\bm P_i^{\ell-}(k)\bm C_i^\top[\bm C_i\bm P_i^{\ell-}(k)\bm C_i^\top+\bm R_i]^{-1},\\
	 \hat {\bm x}_i^\ell(k)&=\hat {\bm x}_i^{\ell-}(k)+\bm K_i^\ell(k)[\bm y_i(k)-\bm C_i\hat {\bm x}_i^{\ell-}(k)],\\
	 \bm P_i^\ell(k)&=[\bm I_{n_i}-\bm K_i^\ell(k)\bm C_i]\bm P_i^{\ell-}(k),
\end{align*}
where $\hat {\bm x}_i^{\ell-}(k)$ and $\hat {\bm x}_i^\ell(k)$ are the \emph{a priori} and the \emph{a posteriori} minimum mean squared error (MMSE) estimates of the state $\bm x_i(k)$ at the $i$-th sensor, $\bm P_i^{\ell-}(k)$ and $\bm P_i^\ell(k)$ are the corresponding estimation error covariances. The recursion starts from $\hat {\bm x}_i^\ell(0)=\bm 0$ and $\bm P_i^\ell(0)=\bm \Pi_i$.

To facilitate the subsequent discussion, we define the Lyapunov and Riccati operators $h_i$ and $g_i:\mathbb{S}_+^{n_i}\mapsto\mathbb{S}_+^{n_i}$ as follows:
\begin{align*}
	h_i(\bm X)&\triangleq \bm A_i\bm X\bm A_i^\top+\bm Q_i,\\
	g_i(\bm X)&\triangleq \bm X-\bm X\bm C_i^\top(\bm C_i\bm X\bm C_i^\top+\bm R_i)^{-1}\bm C_i\bm X.
\end{align*}
Under the detectability and stabilizability assumptions, the estimation
error covariance associated with each local Kalman filter converges exponentially to a steady state from any initial condition~\cite{anderson2012optimal}. Without loss of generality, we assume that the Kalman filter at sensor side has entered steady state, i.e., 
\begin{align*}
	\bm P_i^\ell(k)=\overline {\bm P}_i,\forall k\in\mathbb{N},
\end{align*}
where $\overline {\bm P}_i\succeq\bm 0$ is a unique positive semi-definite solution of discrete algebraic Riccati equation $\bm X=g_i\circ h_i(\bm X)$.

\subsection{Remote Estimator}
We consider a time-slotted communication channel which is shared by $N$ sensors. It is assumed that the remote estimator only has the ability to successfully receive one data packet at each time $k$. In other words, when two or more sensors transmit simultaneously, collision occurs and all the transmitted data packets will be dropped. To schedule the transmission, a time-synchronized transmission scheduler embedded in each sensor makes a decision to determine whether to transmit or not. Specifically, we denote $\theta_i(k):\mathbb{N}\mapsto\{0,1\}$ be the transmission policy of sensor $i$ at time $k$. If the local estimate $\hat {\bm x}_i^\ell(k)$ is scheduled to transmit, $\theta_i(k)=1$; otherwise $\theta_i(k)=0$. Let $\theta(k)\triangleq\{\theta_1(k),\theta_2(k),\ldots,\theta_N(k)\}$, $\theta\triangleq\{\theta(1),\theta(2),\ldots\}$, and $\Theta$ be the set of all feasible schedules. Moreover, we denote $\lambda_i(k)=\prod_{j\neq i,j\in\mathcal{N}}\theta_i(k)[1-\theta_j(k)]$ to indicate the transmission result of sensor $i$ at time $k$. If the local estimate $\bm{\hat x}_i^\ell(k)$ is successfully received at the remote estimator, $\lambda_i(k)=1$; otherwise $\lambda_i(k)=0$. 

As a result, the MMSE state estimate $\hat {\bm x}_i(k)$ and the corresponding estimation error covariance $\bm P_i(k)$ at the remote estimator can be computed by the following recursions:
\begin{align}
\hat{\bm x}_i(k)&=
	\begin{cases}
	\hat {\bm x}_i^\ell(k),&\text{ if }\lambda_i(k)=1,\\
	\bm A_i\hat{\bm x}_i(k-1),\hspace{10pt}&\text{ if }\lambda_i(k)=0,
	\end{cases}\\
\bm P_i(k)&=
	\begin{cases}
	\overline {\bm P}_i,&\text{ if }\lambda_i(k)=1,\\
	h_i(\bm P_i(k-1)),\hspace{5pt}&\text{ if }\lambda_i(k)=0.
	\end{cases}
\end{align}
Note that the estimation error covariance satisfies
\begin{align}
h_i^{t_1}(\overline {\bm P}_i)\succeq h_i^{t_2}(\overline {\bm P}_i)
\end{align}
for any $t_1,t_2\in\mathbb{N}$ with $t_1\geq t_2$~\cite{shi2010kalman}. Furthermore,
\begin{align}
\Tr\left[h_i^{t_1}(\overline {\bm P}_i)\right] \geq \Tr\left[h_i^{t_2}(\overline {\bm P}_i)\right].
\end{align}
This well ordering of the estimation error covariance is helpful for the further analysis.

\subsection{Optimal Sensor Scheduling}
For the considered collision channel, an efficient sensor scheduling scheme plays a crucial role to system estimation performance, which motivates us to consider the following optimization problem:
\begin{align}\label{eqn:optimal-schedule}
	\min_{\theta\in\Theta}\quad&J(\theta)\triangleq\limsup_{K\rightarrow\infty}\frac{1}{K}\sum_{k=0}^{K-1}\sum_{i=1}^{N}\Tr\left[\bm P_i(k)\right]\\
	\rm s.t.\quad& \sum_{i=1}^{N}\lambda_i(k)\leq 1,~\forall k\in\mathbb{N}.\nonumber
\end{align}
The above problem was already studied in~\cite{han2017optimal} and we summarize the main results in the following lemma.

\begin{lemma}\label{lem:property-schedule}
The optimal transmission policy $\theta^\ast$ for the optimization problem in~\eqref{eqn:optimal-schedule} has the following properties:
\begin{enumerate}
	\item Exclusivity: there must be one and only one sensor which transmits at each time $k$;
	\item Periodicity: there exists a period $T\in\mathbb{N}$ such that $\theta_i^\ast(k)=\theta_i^\ast(k+T),\forall i\in\mathcal{N}$;
	\item Uniformity: each sensor must schedule its transmission as uniformly as possible within one period.
\end{enumerate}
\end{lemma}

\begin{IEEEproof}
See Theorem 2 and Theorem 3 in~\cite{han2017optimal}.
\end{IEEEproof}

\subsection{Problem of Interest}
There is no doubt that the minimum average estimation error covariance over an infinite time horizon can be achieved when applying the optimal sensor schedules. However, if there exists a malicious attacker who aims at damaging the estimation quality, i.e., maximizing the objective function $J(\theta)$ by choosing an appropriate attack strategy, the situation will become more involved, which motivates our current work. In this case, the defense strategies that help the system maintain acceptable performance even in the presence of attacks are worth exploring. We will introduce the detailed attack model and propose countermeasures in the following two sections.

\section{Attack Strategy and Performance Analysis}
In this section, we consider the scenario where there exists a malicious attacker who is able to generate random delays on the clocks of the sensors to increase collision, consequently packet dropout, during the transmission. The mathematical formulation and practical implementation of such an attack are discussed. Moreover, the remote estimation performances when the malicious attacker has different system information are investigated and an efficient algorithm is proposed to calculate the optimal attack strategy.

\subsection{Time Synchronization Attack}
To achieve the optimal transmission policy obtained in~\cite{han2017optimal}, GPS timing is applied such that each sensor in the system transmits data packet in a synchronous manner without any collision. We consider a malicious attacker who intentionally disturbs the time synchronization among sensors by injecting arbitrary time delays on sensor clocks. Practically, this can be achieved by introducing a forged GPS signal, as shown in Fig.~\ref{fig:system}. Note that such an attacker does not need to hack into the underlying system or have physical contact to the sensors. It is also difficult to locate the attacker since it can transmit the GPS spoofing signal while moving around the target sensors. These injected time delays may lead to transmission collision, packet dropout and performance degradation. Consequently, it is necessary to analyze attack consequences under different information sets and propose efficient countermeasures, before which some definitions are introduced first.

According to Lemma~\ref{lem:property-schedule}, the optimal transmission policy for each sensor is a periodic $0$-$1$ sequence. Thus, it suffices to investigate the attack effect on sensor's transmission within one period instead of an infinite horizon. To simplify the subsequent discussion, we denote the optimal transmission policy $\bm s_i$ for sensor $i$ within period $T$ as a column vector:
\begin{align}
	\bm s_i\triangleq 
	\begin{bmatrix}
	s_i(0),s_i(1),\ldots,s_i(T-1)
	\end{bmatrix}^\top,
\end{align}
where $s_i(k)\in\{0,1\}$ for all $i\in\mathcal{N}$ and $k\in\{0,1,\ldots,T-1\}$. Let $\bm s\triangleq\{\bm s_1,\bm s_2,\ldots,\bm s_N\}$ be the optimal transmission policy of all sensors. The Hamming weight of sensor $i$ is the number of ones in $\bm s_i$ within a period. The duty factor of sensor $i$ is the fraction of time in which sensor $i$ is transmitting, which equals the Hamming weight divided by the period, i.e.,
\begin{align*}
	f_i\triangleq \frac{1}{T}\sum_{k=0}^{T-1}s_i(k).
\end{align*}
When a random time delay $\tau_i$ is injected on the clock of the $i$-th sensor by the GPS spoofer, the transmission policy of sensor $i$ becomes
\begin{align}
\bm s_i^{(\tau_i)}&\triangleq 
\begin{bmatrix}
s_i^{(\tau_i)}(0),s_i^{(\tau_i)}(1),\ldots,s_i^{(\tau_i)}(T-1)
\end{bmatrix}^\top \nonumber\\
&=
\begin{bmatrix}
s_i(0\oplus\tau_i),s_i(1\oplus\tau_i),\ldots,s_i((T-1)\oplus\tau_i)
\end{bmatrix}^\top,
\end{align}
where $\oplus$ represents addition modulo $T$.

\subsection{Attack without System Knowledge}
In this subsection, we focus on the scenario where the malicious attacker does not have any system knowledge and randomly launches attacks on an arbitrary subset of the sensors. In this regard, we can show that the expected average estimation error covariance of the overall system over an infinite time horizon goes to infinity. This result can be obtained directly from the following theorem.

\begin{theorem}\label{thm:existence_of_attack}
For any optimal transmission policy $\bm s$ with period $T$, there exists an attack strategy under which all the transmitted data packets of sensor $i$ will be dropped if $f_i\leq\frac{1}{2}$.
\end{theorem}

\begin{IEEEproof}
The proof is divided into two parts. The two-sensor scenario is considered first, and as an extension, the proof of the scenario with $N$ sensors is completed.

First, without loss of generality, we assume that $f_1 \leq f_2$ in the two-sensor scenario, which means that the sensor $1$ transmits $T f_1$ times and the sensor $2$ transmits $T f_2$ times in one period $T$. For each transmission policy $\theta$, the resulting average cost $J(\theta)$ in the Problem $1$ can be computed:
\begin{align}
J(\theta)=\frac{1}{T}\bigg\{\sum_{t=0}^{T f_2}a_t \Tr\left[h_1^t\left(\overline {\bm P}_1\right)\right]+\sum_{t=0}^{T f_1}b_t \Tr\left[h_2^t\left(\overline {\bm P}_2\right)\right]\bigg\},
\end{align} 
where $a_t$ and $b_t$ are nonnegative integers satisfying
\begin{align}
\sum_{t=0}^{T f_2}a_t = \sum_{t=0}^{T f_1}b_t =T, \ a_0=T f_1, \ b_0=T f_2,\\
a_0\geq a_1\geq \cdots \geq a_{T f_2}, \ b_0\geq b_1\geq \cdots \geq b_{T f_1}, \label{sensor_philosophy}
\end{align}
$a_0=T f_1$ represents the total number of time slots in a period $T$ when sensor $1$ is scheduled to transmit, and $a_t, t\geq 1$ stands for the total number of time slots that sensor $1$ has continuously idled for $t$ slots after a transmission. Sensor $1$ needs to transmit for exactly $T f_1$ times in a period $T$, and thus the longest waiting time duration during which it cannot transmit is $T-T f_1=T f_2$. The counting number summation for sensor $1$, denoted as $\sum_{t=0}^{T f_2}a_t$, must be $T$. Sensor $2$ has similar situations represented by $b_t$. The case that the sensor $i$ has not been scheduled to transmit for $t+1$ time slots can happen only if it has not been scheduled for $t$ time slots. Therefore, the inequality \eqref{sensor_philosophy} holds. To illustrate the meanings of $a_t$ and $b_t$, we provide a simple example in Fig.~\ref{fig:thm2-1}. A transmission policy in the two-sensor scenario with $T=7, f_1=\frac{2}{7}$ and $f_2=\frac{5}{7}$ is considered. The corresponding $a_0=2, a_1=2, a_2=2, a_3=1, a_4=a_5=0$ and $b_0=5, b_1=2, b_2=0$. 

\begin{figure}[t]
	\centering
	\includegraphics[width=0.35\textwidth]{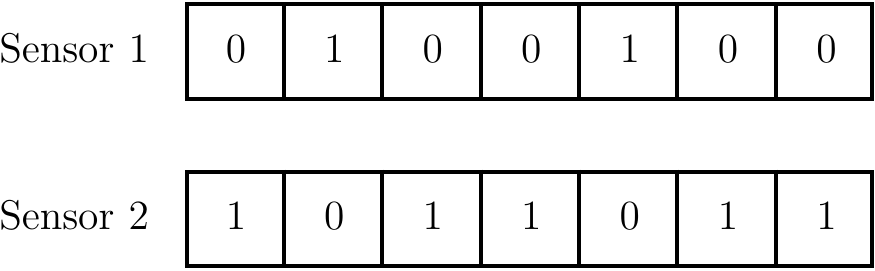}
	\caption{Transmission policy of two sensors with period $T=7$}
	\label{fig:thm2-1}
\end{figure}

Due to the nondecreasing property of $\Tr\left[h_i^{t}\left(\overline {\bm P}_i\right)\right]$ in $t$ and Lemma~\ref{lem:property-schedule}, the average estimation error covariance under the optimal transmission policy $\theta^\ast$ can be obtained as:
\begin{align}
J(\theta^\ast)&=\frac{1}{T}\bigg\{a_0 \Tr\left[h_1^0\left(\overline {\bm P}_1\right)\right]+a_0 \Tr\left[h_1^1\left(\overline {\bm P}_1\right)\right]+\cdots\nonumber\\
&~~~+a_0 \Tr\big[h_1^{\lfloor\frac{T}{a_0}\rfloor-1}\left(\overline {\bm P}_1\right)\big]\nonumber\\
&~~~+\left(T \bmod a_0\right) \Tr\big[h_1^{\lfloor\frac{T}{a_0}\rfloor}\left(\overline {\bm P}_1\right)\big]\nonumber\\
&~~~+b_0\Tr\left[h_2^0\left(\overline {\bm P}_2\right)\right]+\left(T-b_0\right)\Tr\left[h_2^1\left(\overline {\bm P}_2\right)\right]\bigg\}.
\end{align}
In consistency with the \emph{Uniformity} in Lemma~\ref{lem:property-schedule}, the sensor $2$ with the larger duty factor $f_2$ is only allowed to wait for at most one time slot under the optimal scheduling policy. In other words, all the ``$0$'' elements in $\bm s_2$ are isolated by other ``$1$'' elements. According to the \emph{Exclusivity} in Lemma \ref{lem:property-schedule}, all the ``$1$'' elements in $\bm s_1$ are isolated by other ``$0$'' elements. Consequently, the attacker can always set $\tau_2=1$ and construct $\bm s_2^{(1)}$ such that all the ``$1$'' elements in $\bm s_1$ collide with those ``$1$'' elements in $\bm s_2^{(1)}$.

For the general scenario with $N$ sensors, any sensor $i$ with $f_i\leq\frac{1}{2}$ satisfies that ``$1$'' elements in $\bm s_i$ isolated by other ``$0$'' elements according to the \emph{Uniformity}, which can be viewed as sensor 1 in the two-sensor scenario. Then, all the other sensors can be viewed together as sensor 2 in the two-sensor scenario. Similar to the previous case, there always exist a delayed version of sequences $\bm s_1^{(\tau_1)}, \bm s_2^{(\tau_2)}, \ldots, \bm s_{i-1}^{(\tau_{i-1})}, \bm s_{i+1}^{(\tau_{i+1})}, \ldots, \bm s_N^{(\tau_N)}$ with $\tau_j=1,\forall j\in\mathcal{N}\backslash \{i\}$ such that all the transmitted data packets of sensor $i$ are dropped in collisions.
\end{IEEEproof}

\begin{corollary}\label{cor:random-attack}
Consider system~\eqref{eqn:dynamic}--\eqref{eqn:measurement} under optimal transmission policy $\bm s$ with period $T$. When a randomly generated time synchronization attack is launched on the system, the expected average estimation error covariance of the overall system over an infinite time horizon goes to infinity, i.e., 
\begin{align}
	\limsup_{K\rightarrow\infty}\frac{1}{K}\sum_{k=0}^{K-1}\sum_{i=1}^{N}\mathbb{E}\{\Tr\left[\bm P_i(k)\right]\}=\infty.\label{cor:exp_infinity}
\end{align}
\end{corollary}

\begin{IEEEproof}
Theorem \ref{thm:existence_of_attack} shows that for any system under the optimal transmission policy $\bm s$, there always exists an attack strategy such that at least one sensor, e.g., sensor $i$, will never have a chance to successfully transmit its data packet. Correspondingly, the trace of estimation error covariance $\Tr\left[\bm P_i(k)\right]$ diverges to infinity for this unstable process. Note that the attack strategy is randomly chosen from all the finite types of time synchronization attacks. Hence, the equation \eqref{cor:exp_infinity} holds, taking expectation with respect to random attack strategies. 
\end{IEEEproof}

\subsection{Attack with full System Knowledge}
When the malicious attacker has knowledge of all system parameters $\bm A_i$, $\bm C_i$, $\bm Q_i$ and $\bm R_i$, $\forall i\in\mathcal{N}$, it is able to calculate the optimal sensor schedules and launch an attack such that the average estimation error covariance of the overall system goes to infinity, i.e., $\limsup_{K\rightarrow\infty}\frac{1}{K}\sum_{k=0}^{K-1}\sum_{i=1}^{N}\Tr\left[\bm P_i(k)\right]=\infty$. Note that the attacks leading to unbounded estimation error covariance are not unique. Hence, the optimal attack strategy that spoofs the least number of sensors to achieve this goal is worth investigating. In this subsection, we first show that the optimal attack strategy can be solved by the optimization problem summarized in the following theorem and then provide an efficient algorithm to solve this problem. For notation brevity, we denote 
\begin{align*}
	\bm S_i\triangleq[\bm s_i^{(1)},\bm s_i^{(2)},\ldots,\bm s_i^{(T-1)}]\in\{0,1\}^{T\times(T-1)}
\end{align*}
as all possible attacked version of $\bm s_i$,
\begin{align*}
	\bm\varGamma_i\triangleq[\gamma_i^{(1)},\gamma_i^{(2)},\ldots,\gamma_i^{(T-1)}]^\top\in\{0,1\}^{T-1}
\end{align*}
as an indicator vector corresponds to $\bm S_i$, and $\bm e_i$ being $(T-1)(N-1)$ dimensional vector with $N-1$ partitions and all the elements in the $i$-th partition are ones. Moreover, we define
$\bm S_{-i}\triangleq[\bm S_1,\ldots,\bm S_{i-1},\bm S_{i+1},\ldots,\bm S_N]\in\{0,1\}^{T\times(T-1)(N-1)}$, $\bm\varGamma_{-i}\triangleq[\bm\varGamma_1^\top,\ldots,\bm\varGamma_{i-1}^\top,\bm\varGamma_{i+1}^\top,\ldots,\bm\varGamma_N^\top]^\top\in\{0,1\}^{(T-1)(N-1)}$, and $\bm E=[\bm e_1^\top,\bm e_2^\top,\ldots,\bm e_{N-1}^\top]^\top\in\{0,1\}^{(N-1)\times(T-1)(N-1)}$.
 
\begin{theorem}\label{thm:optimal-attack}
Consider system~\eqref{eqn:dynamic}--\eqref{eqn:measurement} under optimal transmission policy $\bm s$ with period $T$. The optimal attack strategy can be obtained by solving the following optimization problem for all $i\in\mathcal{N}$:
\begin{align*}
\mathbf{P_1}:\quad\min_{\bm\varGamma_{-i}}\quad&\Vert\bm\varGamma_{-i}\Vert_1\\
	\rm s.t. \quad&\bm D_i\bm\varGamma_{-i}\preceq_e\bm b_i \\
	&\bm\varGamma_{-i}\in_e\{0,1\},
\end{align*}
where $\bm b_i\triangleq\begin{bmatrix}\bm s_i\\\bm 1\end{bmatrix}\in\{0,1\}^{T+N-1}$, $\bm D_i\triangleq\begin{bmatrix}-\bm S_{-i}\\\bm E\end{bmatrix}\in\{0,1\}^{(T+N-1)\times(T-1)(N-1)}$,   $\Vert\bm x\Vert_1$ stands for $l_1$ norm of $\bm x$, $\preceq_e$ is element-wise inequality, $\bm x\in_e\{0,1\}$ means that each element of $\bm x$ belongs to $\{0,1\}$.
\end{theorem}

\begin{IEEEproof}
According to the definition, an attack strategy is optimal with respect to sensor $i$ if it spoofs the least number of sensors except sensor $i$ such that all the transmitted data packets of sensor $i$ are dropped. Consequently, to obtain the optimal attack strategy with respect to sensor $i$, we need to solve the following optimization problem:
\begin{align*}
\mathbf{P_2}:\quad\min_{\bm\varGamma_{-i}}\quad&\Vert\bm\varGamma_{-i}\Vert_1\\
	\rm s.t. \quad&\bm S_{-i}\bm\varGamma_{-i}-\bm s_i\succeq_e\bm 0\\
	&\bm\varGamma_j\in_e\{0,1\},~\forall j\in\mathcal{N}\backslash \{i\}\\
	&\Vert\bm\varGamma_j\Vert_1\preceq_e\bm 1,~\forall j\in\mathcal{N}\backslash \{i\}.
\end{align*}
Since injecting a time delay $\tau_j$ on sensor $j$, can be represented as the multiplication of $\bm S_j$, the matrix of all possible attacked version of $\bm s_j$, and its corresponding indicator $\bm\varGamma_j$, any attack strategy satisfying the first two constraints guarantees that at least one sensor $j\in\mathcal{N}\backslash \{i\}$ is also transmitting its data packet at the time slot when sensor $i$ is transmitting. Note that only one time delay can be injected on each sensor $j\in\mathcal{N}\backslash \{i\}$, which leads to the third constraint. It can be observed that all the transmitted data packets of sensor $i$ will be dropped under any attack strategy satisfying above three constraints. To obtain the optimal attack, one has to minimize the number of sensors needed to be spoofed, which corresponds to the $l_1$ norm in the objective function.

Combining the first and the third constraints of problem~$\mathbf{P_2}$, one can obtain problem~$\mathbf{P_1}$, which completes the proof.
\end{IEEEproof}

\begin{algorithm}[t]
	\small
	\caption{B\&B Algorithm for Optimal Attack Strategy}
	\label{alg:optimal-attack}
	\begin{algorithmic}[1]
		\State Initialization: $temp_i=+\infty$, $\mathcal N_{live}^i=\mathcal N_{-i}$, $\mathcal N_{del}^i=\emptyset$, $\bm\varGamma_{-i}(\mathcal N_{del}^i)=\emptyset$, $\forall i\in\mathcal{N}$, $val_{opt}=+\infty$, $sol_{opt}=\bm 0$;
		\For {$i=1:N$}
		\State $\bm\varGamma_{-i}^\ast=\BB(\mathcal N_{live}^i,\mathcal N_{del}^i,\bm \varGamma_{-i}(\mathcal N_{del}^i),temp_i)$
		\If{$\Vert\bm\varGamma_{-i}^\ast\Vert_1<val_{opt}$}
		\State $val_{opt}=\Vert\bm\varGamma_{-i}^\ast\Vert_1$
		\State $sol_{opt}=\bm\varGamma_{-i}^\ast$
		\EndIf
		\EndFor
		\Function{$\BB$}{$\mathcal N_{live}^i,\mathcal N_{del}^i,\bm \varGamma_{-i}(\mathcal N_{del}^i),temp_i$}
		\State Solve the following convex optimization problem: 
		\begin{align*}
		\min_{\bm\varGamma_{-i}}\quad&\Vert\bm\varGamma_{-i}\Vert_1\\
		\rm s.t. \quad&~\bm D_i\bm\varGamma_{-i}\preceq_e\bm b_i \\
		&~\bm\varGamma_{-i}(\mathcal N_{live}^i)\in_e[0,1]\\
		&~\bm\varGamma_{-i}(\mathcal N_{-i}\backslash\mathcal N_{live}^i)=\bm\varGamma_{-i}(\mathcal N_{del}^i)
		\end{align*}
		\If{the solution $\bm\varGamma_{-i}^{sol}$ exists and $\Vert\bm\varGamma_{-i}^{sol}\Vert_1<temp_i$}
		\If{the solution $\bm\varGamma_{-i}^{sol}$ is in binary form}
		\State $temp_i=\Vert\bm\varGamma_{-i}^{sol}\Vert_1$
		\State \Return{$\bm\varGamma_{-i}^{sol}$}
		\Else 
		\State Choose $j\in\mathcal N_{live}^i$ and define:
		\begin{align*}
		\hspace{1cm}
		\begin{cases}
		\tilde{\mathcal N}_{live}^i=\mathcal N_{live}^i\backslash\{j\}\\
		\tilde{\mathcal N}_{del}^i=\mathcal N_{del}^i\cup\{j\}\\
		\tilde{\bm\varGamma}_{-i,0}(\mathcal N_{del}^i)=\bm\varGamma_{-i}(\mathcal N_{del}^i)\cup\{\bm\varGamma_j=\bm 0\}\\
		\tilde{\bm\varGamma}_{-i,1}(\mathcal N_{del}^i)=\bm\varGamma_{-i}(\mathcal N_{del}^i)\cup\{\bm\varGamma_j=\bm e_1\}\\
		\qquad\vdots\\
		\tilde{\bm\varGamma}_{-i,T-1}(\mathcal N_{del}^i)=\bm\varGamma_{-i}(\mathcal N_{del}^i)\cup\{\bm\varGamma_j=\bm e_{T-1}\}\\
		\end{cases}
		\end{align*}
		\State Solve the problem for all $k\in\{0,1,\ldots,T-1\}$:
		\begin{align*}
		\hspace{1cm}
		\bm\varGamma_{-i,k}^{sol}=\BB(\tilde{\mathcal N}_{live}^i,\tilde{\mathcal N}_{del}^i,\tilde{\bm\varGamma}_{-i,k}(\mathcal N_{del}^i),temp_i)
		\end{align*}
		\State $\tilde{\bm\varGamma}_{-i}^{sol}=\hspace{-0.3cm}\underset{\{\bm\varGamma_{-i,0}^{sol},\ldots,\bm\varGamma_{-i,T-1}^{sol}\}}{\operatorname{arg\,min}}\hspace{-0.3cm} \{\Vert\bm\varGamma_{-i,0}^{sol}\Vert_1,\ldots,\Vert\bm\varGamma_{-i,T-1}^{sol}\Vert_1\}$
		\If{$\Vert\tilde{\bm\varGamma}_{-i}^{sol}\Vert_1<temp_i$}
		\State $temp_i=\Vert\tilde{\bm\varGamma}_{-i}^{sol}\Vert_1$
		\State \Return $\tilde{\bm\varGamma}_{-i}^{sol}$
		\Else
		\State \Return $null$
		\EndIf
		\EndIf
		\Else
		\State \Return $null$
		\EndIf
		\EndFunction	
	\end{algorithmic}
\end{algorithm}

Due to the binary constraint on $\bm\varGamma_{-i}$, problem~$\mathbf{P_1}$ is a mixed integer programming problem, which cannot be solved by standard convex optimization techniques. Noticing that $\bm\varGamma_{-i}\in_e\{0,1\}$ has $2^{(T-1)(N-1)}$ possible values, a brute-force enumeration for the optimal attack strategy is computationally intractable when the network scale $N$ or the communication period $T$ is large.

Among various algorithms in the literature~\cite{geoffrion1972integer}, Branch-and-Bound (B\&B) algorithm is the most popular one to solve large scale NP-hard combinatorial optimization problems~\cite{clausen1999branch}. Although the algorithm may need to search the entire solution space in the worst case, the use of bounds for the function to be optimized combined with the value of the current best solution enables the algorithm to search a smaller solution space in general. To be specific, we denote $\mathcal N_{live}^i,\mathcal N_{del}^i\subseteq\mathcal N_{-i}\triangleq\mathcal N\backslash\{i\}$ as two sub-index sets of $\mathcal N_{-i}$ for sensor $i$, and $\bm\varGamma_{-i}(\mathcal N_{live}^i),\bm\varGamma_{-i}(\mathcal N_{del}^i)$ as the corresponding collections of elements from $\bm{\varGamma_{-i}}$, respectively. We summarize the B\&B algorithm for optimal attack strategy in Algorithm~\ref{alg:optimal-attack}.

\begin{remark}
	Note that the optimal attack strategy that spoofs the least number of sensors and leads to unbounded estimation error covariance may not be unique. The method summarized in Algorithm~\ref{alg:optimal-attack} only returns one of the optimal attack strategies. If all the optimal attack strategies are needed, one can easily achieve this goal by modifying ``$<$" in lines 11 and 19 to ``$\leq$" and storing all the returned solutions.
\end{remark}

\section{Countermeasure against Time Synchronization Attack}
According to the previous discussion, the optimal transmission policy for each sensor is periodic and synchronized. When a malicious attacker intentionally destroys time synchronization between sensors, nonzero relative offsets are injected and the collided data packets are dropped, which results in degradation of remote estimation performance. To ensure the estimation quality in the presence of time synchronization attacks, we propose a countermeasure based on shift invariant property of the transmission policy in this section. Moreover, we derive the lower and upper bounds of remote estimation error covariance when the proposed defense method is used. 

\subsection{Shift Invariance}
Before proceeding the analysis, we first introduce the definition of shift invariant transmission policy in this subsection.

Let $\mathcal U_N$ be the collection of all ordered tuples of length $1,2,\ldots,N$, whose components are distinct elements in sensor set $\mathcal N$ and sorted in ascending order. It consists of $n$-tuples in the form $(i_1,i_2,\ldots,i_n)$ for some $n$ between $1$ and $N$, and $i_1<i_2<\cdots<i_n$. An element in $\mathcal U_N$ corresponds to an ordered tuple of sensors. For $U=(i_1,i_2,\ldots,i_n)\in\mathcal U_N$ with $i_1<i_2<\cdots<i_n$, the Hamming cross correlation associated with $U$ is defined as
\begin{align}\label{eqn:cross-cor}
	H(\tau_1,\tau_2,\ldots,\tau_n;U)\triangleq\sum_{k=0}^{T-1}\prod_{j=1}^{n}s_{i_j}^{(\tau_j)}(k).
\end{align}
In other words, it counts the number of time slots in a period where all sensors in $U$ transmit simultaneously. When $U$ consists of only one sensor, Hamming cross correlation  reduces to Hamming weight.

A function $F:\{0,1,\ldots,T-1\}^n\mapsto\mathbb N$ is said to be shift invariant if $F(\tau_1,\tau_2,\ldots,\tau_n)$ equals identically to a constant for any choice of $\tau_1,\tau_2,\ldots,\tau_n$. We say a transmission policy set $\{\bm s_1,\bm s_2,\ldots,\bm s_N\}$ is shift invariant if the Hamming cross correlation in~\eqref{eqn:cross-cor} is shift invariant as a function of $\tau_1,\tau_2,\ldots,\tau_n$ for all $U\in\mathcal U_N$.

When the sensors in the ordered tuple $U=(i_1,i_2,\ldots,i_n)\in\mathcal U_N$ are active and the time delay injected by the attacker of sensor $i_j$ is $\tau_j$ for $j=1,2\ldots,n$, the throughput of sensor $i_j$ is defined as
\begin{align}\label{eqn:throughput}
	\theta_j(\tau_1,\tau_2,\ldots,\tau_n;U)\triangleq\frac{1}{T}\sum_{k=0}^{T-1}s_{i_j}^{\tau_j}(k)\prod_{t\neq j}\left(1-s_{i_t}^{\tau_{t}}(k)\right),
\end{align}
where the product is over all $t=1,2,\ldots,n$ except $t=j$. This is the fraction of time slots in which sensor $i_j$ transmits and sensors $i_1,\ldots,i_{j-1},i_{j+1},\ldots,i_n$ keep silent. When $U$ consists of only one sensor, the throughput $\theta(\tau;(i))$ is equivalent to the duty factor $f_i$. A transmission policy set $\{\bm s_1,\bm s_2,\ldots,\bm s_N\}$ is throughput invariant if the throughput in~\eqref{eqn:throughput} is shift invariant as a function for all $U\in\mathcal U_N$ and $j=1,2\ldots,n$. According to Theorem 8 and Theorem 12 in~\cite{zhang2016protocol}, the shift invariance and throughput invariance are equivalent for the transmission channel considered in our work.

\begin{example}
The following is a set of three shift invariant transmission policies with duty factors $f_1=\frac{1}{2}$, $f_2=\frac{1}{2}$, $f_3=\frac{1}{3}$ and period $T=12$:
\begin{align*}
\bm s_1&=\begin{bmatrix}
	0,1,0,1,0,1,0,1,0,1,0,1
	\end{bmatrix}^\top,\\
\bm s_2&=\begin{bmatrix}
	0,0,1,1,0,0,1,1,0,0,1,1
	\end{bmatrix}^\top,\\
\bm s_3&=\begin{bmatrix}
	0,0,0,0,0,0,1,1,1,1,0,0
	\end{bmatrix}^\top.
\end{align*}
Then, we have $H(\tau_1,\tau_2;(1,2))=3$, $H(\tau_1,\tau_3;(1,3))=2$, $H(\tau_2,\tau_3;(2,3))=2$ and $H(\tau_1,\tau_2,\tau_3;(1,2,3))=1$ for all $\tau_1,\tau_2,\tau_3\in\mathbb N$.
\end{example}

\subsection{Performance Bounds}
Note that the shift invariance property introduced in the previous subsection can be used to design countermeasures against time synchronization attack. When shift invariant transmission policies are adopted, each sensor will receive at least one data packet within a period no matter what time delays are injected on the system, which leads to bounded estimation error covariance and improves system robustness. In this subsection, we provide the lower and upper bounds for the remote estimation error covariance when using shift invariant transmission policies, before which we first revisit some useful results obtained in~\cite{shum2009shift}.

\begin{lemma}\label{lem:shift-invariance}
Let $\bm s_1,\bm s_2,\ldots,\bm s_N$ be shift invariant transmission policies for $N$ sensors with duty factors $f_1=\frac{n_1}{d_1},f_2=\frac{n_2}{d_2},\ldots,f_N=\frac{n_N}{d_N}$, respectively, and $\gcd(n_i,d_i)=1,\forall i\in\mathcal N$. Then, the following statements hold:
\begin{enumerate}
	\item The throughput of sensor $i$ is equal to $f_i\prod_{j\neq i}(1-f_j)$;
	\item The period is divisible by and no less than $d_1d_2\cdots d_N$.
\end{enumerate}
\end{lemma}

\begin{IEEEproof}
See Theorem 3 and Theorem 6 in~\cite{shum2009shift}.
\end{IEEEproof}

\begin{theorem}\label{thm:performance-bound}
Consider system~\eqref{eqn:dynamic}--\eqref{eqn:measurement} under shift invariant transmission policies $\bm s_1,\bm s_2,\ldots,\bm s_N$ with duty factors $f_1=\frac{n_1}{d_1},f_2=\frac{n_2}{d_2},\ldots,f_N=\frac{n_N}{d_N}$, $\gcd(n_i,d_i)=1,\forall i\in\mathcal N$, and period $D=d_1d_2\cdots d_N$. When an arbitrary time synchronization attack is launched on the system, the average estimation error covariance of the overall system over an infinite time horizon is lower bounded by
\begin{align} \label{eqn:lower-bound}
\underline {J}(\theta)&=\frac{1}{D}\sum_{i=1}^N\bigg\{N_i\Tr\left[h_i^0(\overline P)\right]+N_i\Tr\left[h_i^1(\overline P)\right]+\cdots\nonumber\\
&~~~+N_i\Tr\big[h_i^{\lfloor\frac{D}{N_i}\rfloor-1}(\overline P)\big]+(D\bmod N_i)\Tr\big[h_i^{\lfloor\frac{D}{N_i}\rfloor}(\overline P)\big]\bigg\}
\end{align}
and upper bounded by
\begin{align} \label{eqn:upper-bound}
\overline {J}(\theta)&=\frac{1}{D}\sum_{i=1}^N\bigg\{N_i\Tr\left[h_i^0(\overline P)\right]+\Tr\left[h_i^1(\overline P)\right] \nonumber\\
&~~~+\Tr\big[h_i^2(\overline P)\big]+\cdots+\Tr\big[h_i^{(D-N_i)}(\overline P)\big]\bigg\},
\end{align}
where $N_i=n_i\prod_{j\neq i}(d_j-n_j)$ for all $i\in\mathcal N$.
\end{theorem}

\begin{IEEEproof}
According to Lemma~\ref{lem:shift-invariance}, when sensor $i$ adopts shift invariant transmission policy $\bm s_i$ with duty factor $f_i=\frac{n_i}{d_i}$ and period $D=d_1d_2\cdots d_N$, it will receive $N_i=Df_i\prod_{j\neq i}(1-f_j)=n_i\prod_{j\neq i}(d_j-n_j)$ data packets within one period under arbitrary time synchronization attack.

Consequently, the lower bound of the average estimation error covariance over an infinite time horizon of the overall system is achieved when all the data packets are uniformly received, which corresponds to~\eqref{eqn:lower-bound}. The upper bound is achieved when all the data packets are consecutively received, which is consistent with~\eqref{eqn:upper-bound}.
\end{IEEEproof}

\subsection{Construction Method}
In this subsection, we describe how to construct the shift invariant transmission policies with duty factor $0<f_i=\frac{n_i}{d_i}<1$ and $\gcd(n_i,d_i)=1,\forall i\in\mathcal N$. 

In our construction, the $i$-th transmission policy has period $D_{i}\triangleq\prod_{j=1}^i d_i$ and $D_N=d_1d_2\cdots d_N=D$ is the common period for the whole transmission policy set. For convenience, $D_0$ is defined to be $1$. Then, the shift invariant transmission policy for sensor $i$, $i\in\mathcal N$, is constructed as follows. Select $D_{i-1}$ vectors of length $d_i$, say $\bm\sigma_{i1},\bm\sigma_{i2},\ldots,\bm\sigma_{iD_{i-1}}$, such that the Hamming weights of them are all equal to $n_i$, and interleave these $D_{i-1}$ vectors in the following manner:
\begin{align*}
\left[\right.&\bm\sigma_{i1}(0),\bm\sigma_{i2}(0),\ldots,\bm\sigma_{iD_{i-1}}(0),\\
	  &\bm\sigma_{i1}(1),\bm\sigma_{i2}(1),\ldots,\bm\sigma_{iD_{i-1}}(1),\ldots,\\
	  &\bm\sigma_{i1}(d_i-1),\bm\sigma_{i2}(d_i-1),\ldots,\bm\sigma_{iD_{i-1}}(d_i-1)\left.\right]^\top.
\end{align*} 

\begin{example} 
For $N=2$ with duty factors $f_1=\frac{1}{4}$ and $f_2=\frac{1}{3}$, we pick $\bm\sigma_{11}=[0,0,0,1]^\top$, $\bm\sigma_{21}=[0,0,1]^\top$, $\bm\sigma_{22}=[0,1,0]^\top$ and $\bm\sigma_{23}=\bm\sigma_{24}=[1,0,0]^\top$. The two constructed shift invariant transmission policies are
\begin{align*}
\bm s_1&=\begin{bmatrix}
		0,0,0,1,0,0,0,1,0,0,0,1
		\end{bmatrix}^\top,\\
\bm s_2&=\begin{bmatrix}
		0,0,1,1,0,1,0,0,1,0,0,0
		\end{bmatrix}^\top
\end{align*}
with period $T=12$. The first policy is obtained by repeating $[0,0,0,1]^\top$ periodically. The second policy is obtained by reading out the rows from top to bottom of matrix
\begin{align*}
[\bm\sigma_{21},\bm\sigma_{22},\bm\sigma_{23},\bm\sigma_{24}]=
	\begin{bmatrix}
		0& 0& 1& 1\\
		0& 1& 0& 0\\
		1& 0& 0& 0
	\end{bmatrix}.
\end{align*}
\end{example}

Observed from Example 2, different shift invariant transmission policies can be obtained by choosing different duty factors and $\bm\sigma_{ij}$, where $i\in\mathcal{N}, j=1,2,\ldots,D_{i-1}$. Two heuristic construction methods are considered in this work: one is to construct shift invariant transmission policies that preserve the same duty factors of the optimal transmission policies without attacks; the other is to construct shift invariant transmission policies that achieve the shortest period. Note that the shortest period $2^N$ is achieved when the duty factor is $f_i=\frac{1}{2}$ for all $i\in\mathcal{N}$. Particularly, in this case, each sensor receives exactly one packet within a period, and thus the lower bound $\underline J(\theta)$ coincides with the upper bound $\overline J(\theta)$. These two construction methods will be compared through simulation examples in the next section.

\section{Simulation Example}
In this section, we provide some numerical examples to illustrate the main theoretical results. We consider a scenario where three sensors monitor three different dynamic processes. The system parameters are given as follows:
\begin{align*}
\bm A_1&=\begin{bmatrix}
		1.01 &0.5\\
		0 &0.2
		\end{bmatrix},
\bm A_2=\begin{bmatrix}
		1.02 &0.4\\
		0 &0.15
		\end{bmatrix},
\bm A_3=\begin{bmatrix}
		1.03 &0.6\\
		0 &0.1
		\end{bmatrix},\\
\bm C_1&=\begin{bmatrix}
		1 &1
		\end{bmatrix},
\bm C_2=\begin{bmatrix}
		1 &1
		\end{bmatrix},
\bm C_3=\begin{bmatrix}
		1 &1
		\end{bmatrix},\\
\bm Q_1&=\begin{bmatrix}
		0.2 &0\\
		0 &0.2
		\end{bmatrix},
\bm Q_2=\begin{bmatrix}
		0.1 &0\\
		0 &0.15
		\end{bmatrix},
\bm Q_3=\begin{bmatrix}
		0.1 &0\\
		0 &0.2
		\end{bmatrix},\\
\bm R_1&=1,
\bm R_2=1,
\bm R_3=1.				
\end{align*}

By solving the optimization problem in~\eqref{eqn:optimal-schedule}, the optimal transmission policy for each sensor is obtained as
\begin{align*}
\bm s_1&=\begin{bmatrix}
	0,0,1
	\end{bmatrix}^\top,~~
\bm s_2=\begin{bmatrix}
	0,1,0
	\end{bmatrix}^\top,~~
\bm s_3=\begin{bmatrix}
	1,0,0
	\end{bmatrix}^\top,
\end{align*}
with duty factor $f_1=\frac{1}{3}$, $f_2=\frac{1}{3}$, $f_3=\frac{1}{3}$, respectively, and period $T=3$. Then, by solving problem~$\mathbf{P_1}$ using the proposed B\&B algorithm, the optimal attack strategy is obtained as $\tau_1=0,\tau_2=0,\tau_3=2$.

\begin{figure}[t]
	\centering
	\includegraphics[width=0.48\textwidth]{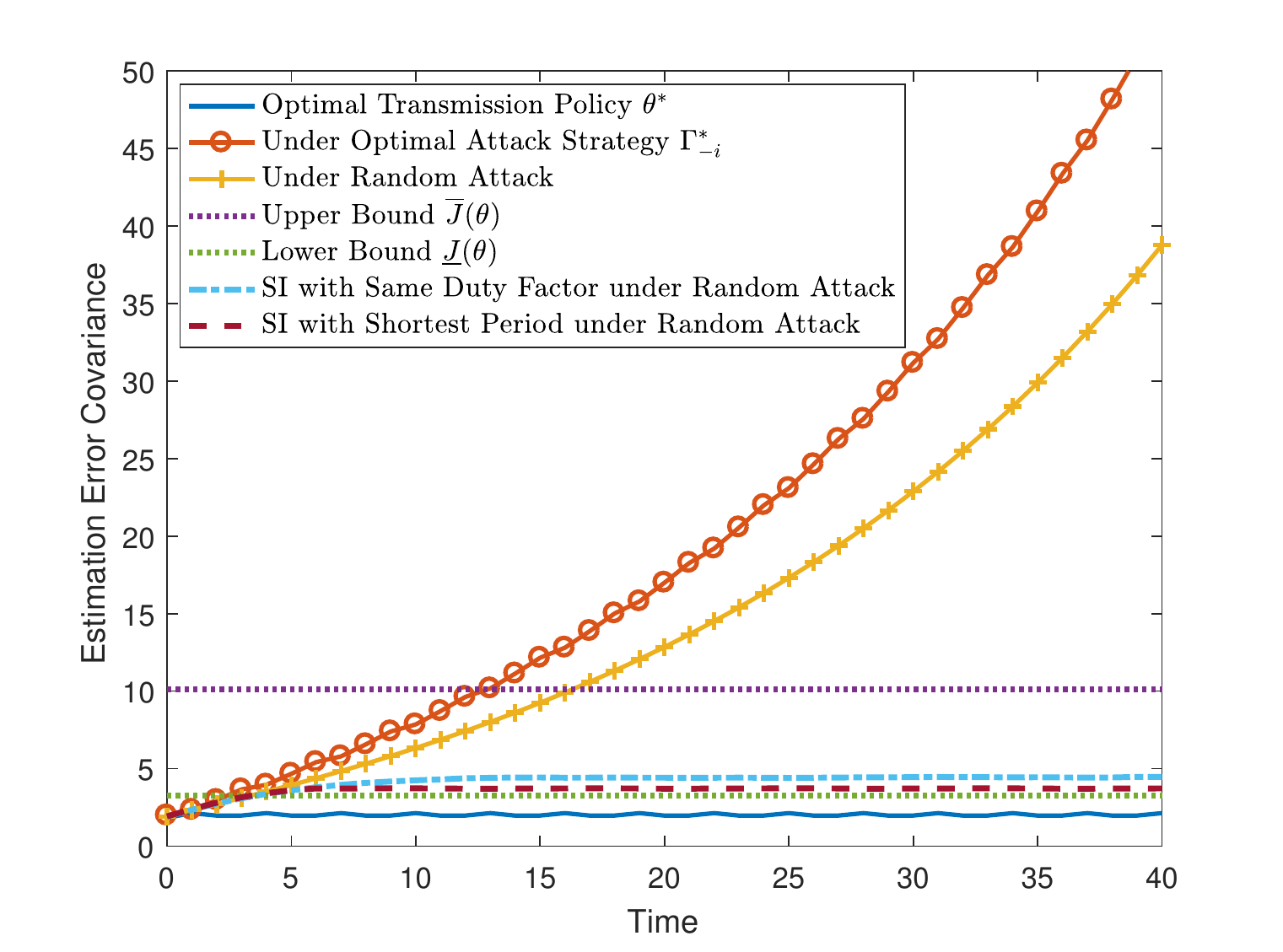}
	\caption{Remote estimation performance under different scenarios.}
	\label{fig:error-cov}
\end{figure}

The infinite-horizon average estimation error covariance of the overall system under different scenarios are shown in Fig.~\ref{fig:error-cov}. The blue solid line represents the estimation error covariance under the optimal transmission policy without attack. The red circle line and the yellow plus line correspond to the estimation performances under the optimal and randomly generated time synchronization attacks, respectively. The blue dash-dot line stands for the expected estimation error covariance when using shift invariant transmission policy that preserves the same duty factors of optimal one under random attack, i.e., $f_1=\frac{1}{3}$, $f_2=\frac{1}{3}$, $f_3=\frac{1}{3}$, $T=27$. Here the expectation is taken with respect to the different choices of $\bm\sigma_{ij}, i\in\mathcal{N}, j=1,2,\ldots,D_{i-1}$ and the randomly generated attacks. The purple dotted line and the green dotted line are the corresponding performance bounds derived in Theorem~\ref{thm:performance-bound}. The red dashed line stands for the estimation error covariance when using shift invariant transmission policy that achieves the shortest period, i.e., $f_1=\frac{1}{2}$, $f_2=\frac{1}{2}$, $f_3=\frac{1}{2}$, $T=8$. It can be observed that the remote estimation error covariance diverges exponentially fast when the system is under the optimal or the randomly generated time synchronization attack, which is consistent with the results obtained in Section~III. On the other hand, the remote estimation error covariance is bounded when shift invariant transmission policies are adopted, which demonstrates the effectiveness of the proposed countermeasure. Moreover, the shift invariant transmission policy with the shortest period achieves a better performance compared to that with the same duty factor.

\section{Conclusion}
In this paper, we studied time synchronization attack against multi-system scheduling in a remote state estimation scenario. For the case that the attacker does not have any system knowledge, we showed that it is able to make the expected average estimation error covariance of the overall system go to infinity. For the case that the attacker has full system knowledge, we proposed an efficient algorithm to solve the optimal attack that spoofs the least number of sensors and leads to unbounded average estimation error covariance of the overall system. To mitigate the attack consequence, we further proposed countermeasures and characterized the lower and upper bounds for system estimation performance when using shift invariant transmission policies. Simulation and comparison were provided to demonstrate the analytical results. 

For the future work, one possible direction is to investigate optimal sensor schedule, optimal time synchronization attack and countermeasures when multiple packet reception is allowed. Additionally, it would be interesting to analyze the explicit estimation performance when different shift invariant transmission policies are adopted.
	
\bibliographystyle{IEEETran}
%\bibliography{reference}

\end{document}